\documentstyle[prb,aps]{revtex}
\begin{document}
\input psfig.tex
\twocolumn
\draft
\def\half{{1\over 2}}
\def\tr{{\rm tr~}}

\preprint{RU-94-94 \\  OUTP-97-22S}
\title{ Chiral  Liquids in 1-d: A New Class of NFL-Fixed Points }
\author{Natan Andrei$^{1}$, Michael R. Douglas$^{1}$ and  Andr\'es
 Jerez$^{1,2}$\cite{ill}}
\address{$^{1}$ Department of Physics and Astronomy, Rutgers University,
Piscataway, NJ 08855}
\address{$^{2}$ University of Oxford, Department of Physics, Theoretical
Physics, 1 Keble Road, \\  Oxford OX1 3NP, United Kingdom \\}

\date{\today}

\maketitle

\begin{abstract}

We identify a  non Fermi Liquid (NFL) class of fixed points describing
the infrared behaviour of interacting chiral fermions in one dimension. The
thermodynamic properties and asymptotic correlation functions are
characterized by universal exponents, which we determine by means of
conformal and Bethe-Ansatz techniques. The mechanism leading to the 
breakdown of Fermi-Liquid theory is quite general and can be expected to
be realized in systems with broken T-invariance.
As an example  we study the edge states of
interacting QHE systems. We calculate
the universal frequency dependence of the spin conductance in
these systems as well as the NMR response.

\end{abstract}

\pacs{71.27.+a}


Models displaying Non-Fermi-Liquid (NFL) behavior have been intensely
studied following the suggestion that the normal phase of
the  cuprate superconductors cannot be described in
 terms of the conventional Landau
theory \cite{pwa}.
A much studied example are the Luttinger
liquids, a class of one dimensional models  whose 
behavior in the infrared (IR) limit is given by the Luttinger model
(in the language
of conformal field theory a $c=1$ bosonic field theory)  characterized  
by single particle
correlation functions having only cuts with non-universal exponents.

In this article we identify a large new class of IR fixed points,  to
be
 refered to as  {\it chiral fluids},
 available  in one dimension to
 interacting fermions with different numbers of left and right moving
degrees of freedom.
The new fixed points  describe NFL physics, and are  universal
in character, independent of the strength
 of the interaction.

\begin{figure}[here]
\psfig{file=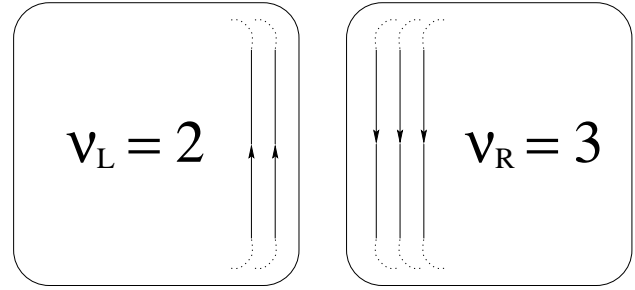,width=3.250in}
\bigskip
\caption{Possible realization of chiral liquids. Two integer QHE
systems side by side with filling factors $\nu_L=f_L=2$, and
$\nu_R=f_R=3$. The lines represent the edge states.}
\label{fig1}
\end{figure}

Systems described by  chiral fluids 
break T-invariance; this may occur explicitly, in the presence of a magnetic
 field, or spontaneously by
the interactions.
The edge states in a paired 
sample of integer QHE systems furnish a concrete example:
Thus consider two integer QHE system arranged side by side (see Fig.
\ref{fig1}),
with the left (right) filling factor $\nu_L=f_L$ ($\nu_R=f_R$)
with $f_L$ ($f_R$) integers. The samples can be considered unpolarized
over a large range of the parameter space due to the smallness of the
Zeeman energy  \cite{eisen}, $E_z\sim 0.4 K$. 
Along the common edge there are then $f_L$
channels of left moving and $f_R$ channels of right moving fermions.
These are good quantum numbers even for a non-vanishing overlap
integral $t$ since only virtual hoppings are allowed between the left and right
samples which are gapped by the cyclotron frequency $\omega_c$. 
Below we shall  see that the resulting interaction leads to some novel
features in the low energy dynamics.

Fixed points of one dimensional quantum field theories 
 can always be related to two dimensional
statistical models, and thus the general theory of such fixed points is
conformal field theory (CFT)~\cite{MS}.
A general classification of fixed points (likely to arise physically)
 is known,
as are calculational techniques for determining  arbitrary Green's functions.
Various approaches to the subject exist; one can study an action which flows
to the fixed point of interest, or define a model by its symmetries and use
techniques of `current algebra' to compute correlation functions.
We will do both in this note.  A direct way to search for and classify NFL
behavior is to use algebraic results for correlation functions.
However, merely showing that a certain fixed point exists abstractly
does not make it a physical model.
The main points of our work are first, that we have explicit models
of interacting fermions which realize these fixed points,
and second, that they
illustrate a very general mechanism for producing NFL behavior.

We start with the 
hamiltonian,
\begin{eqnarray}
H_0 &=& -i v_F \int dx \left(\sum
_{r=1}^{f_R}\psi_{R,a,r}^*(x)\partial_x 
\psi_{R,a,r}(x)  \right. \nonumber \\ && - \left.
\sum_{l=1}^{f_L}\psi_{L,a,l}^*(x)\partial_x \psi_{L,a,l}(x)\right).
\label{freef}
\end{eqnarray}
The fields $\psi^*_{R,a,r}(x)$ ($\psi^*_{L,a,l}(x)$),  with
 $a=\pm 1$  the spin
index, and   $r=1...f_{R},(l=1...f_{L})$ the right (left) flavor index,
 create
right-(left-) moving particles with a linearized dispersion
$\epsilon = \pm v_F (k-k_F)$.
Since the flavor and spin indices enter the hamiltonian
on  equal footing, it has the symmetry $U(2f_L)\otimes U(2f_R)$
under independent unitary transformations on $\psi_L$ and $\psi_R$.

The first interacting model we consider, to illustrate some of the
ideas, is the familiar example of
the Luttinger model obtained by adding interactions in the charge sector. The
hamiltonian is (here we choose $f_R=f_L$),
\begin{eqnarray}
H&=&H_0 + g_{2c} \int dx \rho_L(x)\rho_R(x) \nonumber \\ &+&
g_{4c} \int dx \left( \rho_L(x)\rho_L(x)+ \rho_R(x)\rho_R(x)\right),
\label{luttint}
\end{eqnarray}
where the right-moving charge-density is
 $\rho_R(x)=\sum _{r=1}^{f_R}\psi_{R,a,r}^*(x)\psi_{R,a,r}(x)$, and a
similar definition holds for the left movers.
Renormalization group calculations \cite{sol} indicate that
 the model is conformally invariant (in the
infinite cut-off limit.)
As is well known, upon expressing the hamiltonian in terms of bosonic
fields it becomes quadratic and can be solved by a Bogoliubov rotation.
 The $g_{4c}$ term modifies the  charge velocity (without
destroying the FL property at the Fermi surface \cite{voit}),
while the effect of the $g_{2c}$ term is to modify the exponents of the
fermionic correlation functions destroying the pole structure
characteristic of a FL \cite{hald}.

In the language of CFT, the essential features of this model which make
it a NFL are as follows.
First, the model can be decomposed into two non-interacting sectors,
a spin/flavor sector and a charge sector.  The decomposition is
\begin{equation}\label{lutt}
SU(2f_L)_1\times U(1) \otimes
SU(2f_R)_1\times U(1)
\end{equation}
in a notation we proceed to describe.
The notation ``$SU(N)_k$'' stands for a level-$k$ chiral WZW model \cite{witten}.
It is the unique chiral CFT containing
$SU(N)$ current algebra of central charge $k$ and no other degrees
of freedom. The `central charge' $k$
determines the two-point function of the symmetry currents,
$<J^a(x_1) J^b(x_2)>=k/(x_1-x_2)^2$.
We normalize the currents so that they have the same commutation relations
as generators of the fundamental representation
$t^a$ satisfying $\tr t^a t^b=\half\delta^{ab}$
(for $SU(2)$, $t^a=\half\sigma^a$).
Then, $f$ flavors of complex fermions have $SU(f)_{k=1}$ current algebra.
In a CFT the currents and stress tensor are chiral, so
there are independent left and right central charges $k_L$ and $k_R$.

A similar quantity is the conformal central charge $c$, associated with the
two-point function of the stress tensor.  We will always use the word
`conformal' when referring to it. Here it takes the value,
\begin{eqnarray}
c= \frac{(N^2-1)k}{(k+N)}.
\label{central}
\end{eqnarray}

General results of CFT
show that the IR limit of the specific heat of any gapless model
will be linear, $c_V= {\pi \over 12}(c_L+c_R) T$.
Similarly the IR limit of the
magnetic susceptibility of a spin system is the constant
$\chi=(k_L+k_R) \nu_0$,
($\nu_0= 1/ \pi v_F $ the density of states per unit length)
whether the fixed point describes a FL or not.

The ``$U(1)$''  also denotes  a chiral CFT: it
 contains the $U(1)$ current algebra and has $c=1$.
We have indicated by ``$\times$''  a product of theories which are 
essentially decoupled;  ``$\otimes$'' has the same meaning but separates 
the left and
right moving subsectors.

To prove that this product of decoupled theories is equivalent to the
Fermi theory 
(\ref{freef}), one first observes that the Fermi theory
has the stated current algebra,
showing that it must at least contain
this product of chiral WZW models.  One then computes the sum
of conformal central charges of (\ref{lutt}) and observes that it is equal
to that of the Fermi theory.  Thus any degrees of freedom not accounted for
would form a unitary CFT of central charge zero, which is trivial.
The (right-moving) fermion creation
operator is a product of operators from the two sectors,
$\psi_{R,a,r} = e^{i\phi_R}\ g^{ar}_{R}$,
where $g^{ar}$ is a field in the spin/flavor
sector, transforming in the fundamental representation of
$SU(2f_R)$, and $\phi_R$ is the
charge field.

The $U(1)\otimes U(1)$ charge sector
contains a marginal operator which changes the central charge.
As a result 
the dimension of $e^{i\phi}$ will change, and and so
 will  dimension of
$\psi$ from $1/2$ to (say) $\Delta$.
One must show by explicit computation that $g_{2c}$ couples to this operator
(it clearly does not affect the spin/flavor sector).
The reward for doing this is that since there is no scale at the IR fixed
point, the IR limit of the
single particle Green's function is uniquely determined to be
\begin{equation}
<\psi(x_1)\psi^*(x_2)> = (x_1-x_2)^{-2\Delta}
\end{equation}
and $\Delta\ne\half$ directly implies NFL behavior.

It is this strategy which we will generalize to find new fixed points.
The kinematics of one dimension permits many
decompositions of the sort we just used, because
a linear combination of any number of  left (right)-movers is again a
left (right)-mover.
A powerful way to find them is to choose
other decompositions of the symmetry group
of the theory.  One can then build general states and operators using the
action of the symmetry currents on a finite set of primary fields.

We will discuss models with interactions between the spin-densities
(currents, in the CFT language),
$S_L^i$ and $S_R^i$,
where $S^i_L(x)=\sum_l S^i_{L,l}(x)\equiv \sum_{a,b,l}\psi^*_{L,a,l}
\sigma^i_{a,b} \psi_{L,b,l}
$ (similarly for $S^i_R$).
These satisfy two commuting
$SU(2)$ current algebras (Kac-Moody algebras) with
central charges $f_L$ and $f_R$.

Now the appropriate decomposition to make is into a product of three
sectors, containing the spin, flavor, and charge symmetries \cite{al}.
The free fermion theory (\ref{freef}) is equivalent to the model
\begin{equation}\label{bosonize}
SU(2)_{f_L} \!\! \times \! SU(f_L)_2 \! \times \! U(1) \otimes
SU(2)_{f_R} \! \! \times \! SU(f_R)_2 \! \times \!  U(1)
\end{equation}
One proves this the same way as for (\ref{lutt}), by comparing the
symmetries and central charges.
The currents spanning the left-spin 
$SU(2)_{f_L}$
theory are $S^i_L(x)$, and the flavor currents spanning the
left-flavor $SU(f_L)_2$ theory are
$F^A_L(x)=\sum_a F^A_{L,a}(x)\equiv \sum_{a,l}\psi^*_{L,a,l}
\lambda^A_{l,k} \psi_{L,a,k}$ with $\lambda^A$ in the fundamental
 representation of $SU(f_L)$. Similar comments apply to the right hand sector.
The fermion creation operator now becomes
$
\psi_{R,a, r}(x) = g_R^a(x)~ h_R^r(x)~e^{i\phi_R(x)}
$
where $g^a$ and $h^r$ are  fields in the spin and flavor
sectors, respectively, transforming in the
fundamental representation of
$SU(2)$ and $SU(f_R)$, and $\phi_R$ is the
charge field.
The dynamics of the three sectors decouples.

\bigskip

We proceed  to study IR fixed points
of models obtained by
adding spin-exchange interactions to $H_0$,
\begin{equation}
H=H_0+\sum_{r,l,i}  \int dx \; (g_s)^{rl}S^i_{R,r}(x) S^i_{L,l}(x)\; .
\label{eq}
\end{equation}
with $(g_s)$ a matrix of couplings. The models are isotropic when the
matrix elements are all equal, $(g_s)^{rl}=g_s$.
The interaction breaks the separate left and right
$SU(2)$ spin symmetries but preserves a global $SU(2)$.

Standard perturbative calculations show that this perturbation
is marginally relevant and
destabilizes the weak coupling fixed point.
We will first give general arguments from CFT to identify the
resulting IR fixed point.
We will then describe thermodynamic Bethe Ansatz (TBA)
results which confirm this identification and allow studying the flow
in detail.

We begin by discussing the flavor-isotropic
model, characterized by one coupling $g_s$.
For $f_R$=$f_L$, the model is chirally invariant
and  flows to a strong coupling fixed point generating a
mass gap.  If we use the decomposition (\ref{bosonize}),
the interaction will affect only the spin sector, and thus the IR fixed
point will still be non-trivial, describing gapless flavor and charge
excitations.  However the single particle Green's function necessarily couples
to the spin sector as well and thus sees the gap.

We now argue that for $f_R>f_L$,
the spin sector of the model is gapless, and flows
to a non-trivial fixed point.
We use the following result: under any flow,
the difference of conformal central
charges (here referring to the spin sector only) 
between left and right, $c_R-c_L$,
and the difference of Kac-Moody central charges, $f_R-f_L$,
must be preserved.
The simplest proof uses the concept of `anomaly matching,'
familiar in the particle physics literature \cite{thoof}.
Let us make the argument for the difference of the Kac-Moody central
charges $f_R-f_L$.  This is the coefficient of the two-point function
of the  spin current, and if it is non-zero
this current cannot be consistently coupled to an $SU(2)$ gauge field.
The theory can be made consistent  by adding 
decoupled chiral degrees of freedom to bring the total Kac-Moody
charge to zero.
Once we cancel the anomaly, whatever theory we flow to in the IR can manifestly
be coupled to a gauge field.
Since the extra chiral degrees of freedom do not participate in the flow,
the original $f_R-f_L$ must be preserved.
The same argument can be made for $c_R-c_L$ by considering
the coupling of the stress-tensor to a two-dimensional metric.

Since $c_R-c_L\ne 0$ the IR fixed point is non-trivial,
and since $f_R-f_L>0$ it has a non-trivial $SU(2)$ spin symmetry.
On general grounds one expects to get the theory with the lowest
possible conformal central charge $c_R+c_L$, consistent with the known
$c_R-c_L$, $f_R-f_L$ and the symmetries.  This is because
$c=c_R+c_L$ always decreases under RG flow
(the $c$-theorem \cite{Zamolodchikov}) and thus
a generic perturbation will send us to this theory.

In fact there is a unique such fixed point theory.
It is
\begin{equation}\label{coset-fp}
{SU(2)_{f_L}\times
 SU(2)_{f_R-f_L}\over SU(2)_{f_R}} \otimes SU(2)_{f_R-f_L}. 
\label{eq7}
\end{equation}
The right movers are the chiral WZW model
$SU(2)_{f_R-f_L}$ while
on the left, we introduced a new notation for `quotient,'
which denotes a chiral coset CFT ~\cite{cft}.
This sector is spin singlet and has conformal central charge which
is the difference of those of the numerator and denominator,
so this ansatz matches $c_R-c_L$.

We will confirm this prediction of the IR fixed point
for the model (\ref{eq}) below by appealing to TBA results.
It is important to realize however that the result is more general
than these specific models and follows from the two assumptions
of chirality and non-abelian (in our case $SU(2)$ spin) symmetry.
(Although the same argument can be made for $U(1)$,
the fixed points one then obtains are not really new.)
We refer to this mechanism as `chiral stabilization.'

The UV and IR limits of the specific heat and the magnetic susceptibility
can be determined from the central charges.
The specific heat will be linear in $T$
in the UV and in the IR
(with corrections \cite{corr}), and will undergo the flow:
\begin{eqnarray}
\lefteqn{c_V^{uv}={\pi \over 6} ( f_L+f_R)T} && \nonumber \\ && \;
\longrightarrow\; 
c_V^{ir}={\pi \over 6} \left(f_L+f_R +{3(f_R-f_L) \over f_R-f_L +2}
-{3f_R \over f_R +2} \right)T  
\label{estodo}
\end{eqnarray}
where we also included charge and flavor contributions.
The flow in the susceptibility will be,
 $\chi^{uv}=(f_R+f_L) \nu_0 \rightarrow \chi^{ir}= (f_R-f_L) \nu_0$,
leading to a Wilson ratio $R_W=\left(f_L+f_R +{3(f_R-f_L)
\over f_R-f_L +2}
-{3f_R \over f_R +2} \right)/(f_R-f_L)$.

We discuss now the operators around the IR fixed point. A $SU(2)_k$ theory
contains primary fields $\phi^{jm,\bar{j}\bar{m}}(x,t)$
transforming under a particular left and right representation of the
symmetry. There is a finite number of operators allowed: $0 \le
j,\bar{j} \le k/2$, and their dimension is $h = {j(j+1) \over k+2}$.
For the coset theory   there is a single primary
$\phi^{j,j'}_{j''}$ for each choice $0\le j\le f_L/2$,
$0\le j'\le (f_R-f_L)/2$ and $0\le j''\le f_R/2$.
The dimension of the primary is the difference of the dimensions of
the group primaries, up to an integer. We can thus match the physical
fields with the operator basis around the fixed point and  read off
the IR behavior of the correlation functions,
\begin{eqnarray*}
&<& \psi^*_{L,a,l}(x,t)\psi_{L,a',l'}(0,0)>  \\ &&
\rightarrow  \delta^{a a'}
\delta^{l l'}
(x-v_Ft)^{-(1+\delta_L )}
(x+v_Ft)^{-\delta_L } \nonumber \\
&<& \psi^*_{R,a,r}(x,t)\psi_{R,a',r'}(0,0)> \\ && 
\rightarrow
\delta^{a a'}
\delta^{rr'} (x-v_Ft)^{-\delta_R }(x+v_Ft)^{-(1+\delta_R)}  \\
&< & \!\! S^i_L(x, t) S^j_L(0, 0)> \\ && \rightarrow
\delta^{ij} (x+v_Ft)^{-2}
 (x^2-v^2_Ft^2)^{-4/(f_R -f_L+2)}
\nonumber \\
&<& \!\! S^i_R(x, t) S^j_R(0,0)> \rightarrow \delta^{ij}
 (x-v_Ft)^{-2}
\end{eqnarray*}
with $\delta_L=3/2(f_R-f_L+2)$ and $\delta_R=3f_L
/2(f_R-f_L+2)(f_R-f_L)$. This holds for any value of $f_L < f_R$ 
except  when $f_R-f_L=1$, in which case the anomalous dimension in
 the $S_L$ correlation function is 2.

\begin{figure}[here]
\psfig{file=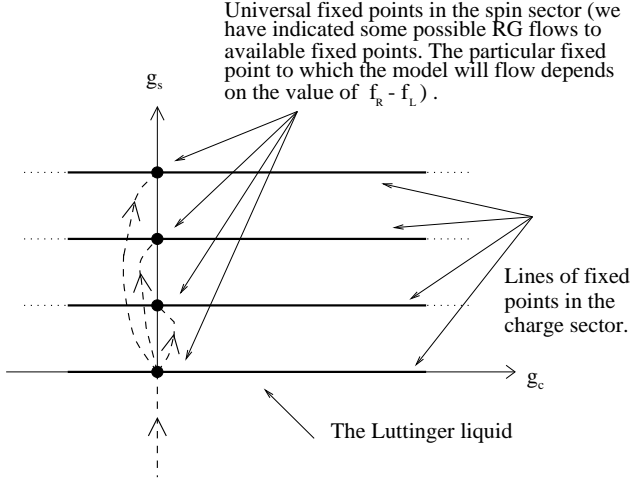,width=3.250in}
\bigskip
\caption{Fixed point structure of the model.}
\label{fig2}
\end{figure}

These results establish that the model (\ref{eq}) is a NFL.
The momentum distributions for small momenta are
$ n_{\alpha}(k)   \sim |k-k_F|^{2\delta_\alpha}$.
In these expressions the left and right components move with velocity
$v_F$.   One can include as well interactions in the charge degrees of freedom
 of the form (\ref{luttint}).
The coupling $g_{4c}$ would modify $v_F$, while the coupling $g_{2c}$ would
again vary the central charge in the $U(1)$ charge sector.  Thus each
of our models can flow to a line of fixed points, generalizing the
Luttinger liquid. See Fig. \ref{fig2}.

These fixed points are robust against small exchange asymmetry, given
by perturbations of the form $t_{ij}S^i_RS^j_L$. The IR limit
 of the perturbation
is expected to contain the operator of lowest dimension with the
same quantum numbers.  For $SU(2)$, the symmetry breaking operators have
a spin 1 ($t_{ij}=-t_{ji}$) component and a spin 2 ($t_{ij}=t_{ji}$) component.
These will flow to operators with the $SU(2)$ quantum numbers realized
entirely on the right (for $f_R>f_L$).

The simplest case to analyze is $f_R-f_L=1$, because the chiral WZW
model appearing in (\ref{coset-fp}) has no primary fields of spin $j>\half$,
and thus the leading operators with $SU(2)$ spins $1$ and $2$ are $f^{ijk}J^k$
and $:J^iJ^j:$.  The perturbation is a scalar operator and if
there also exists an operator in the coset sector with the same dimension,
their product will be the IR limit.
The stress-tensor is dimension $2$ and thus the spin $2$ perturbation flows
to $:J^iJ^j:\bar T$, an irrelevant operator.
It is a non-trivial fact that the cosets appearing in (\ref{coset-fp})
(the BPZ minimal models) do not contain dimension 1 operators, and thus the
spin $1$ perturbation must flow
to an operator of higher dimension, also irrelevant.

Given both the UV and IR results, it is very plausible that small symmetry
breaking perturbations are irrelevant all along the flow.
For $f_R-f_L>1$, primary fields of $SU(2)$ spin $1$ will exist, and then
the IR limit of the symmetry breaking perturbation can be relevant.

By contrast, flavor anisotropy tends to be relevant.
Here we discuss the various limits of extreme anisotropy.
These can be modeled as a sequence of flows, each
of the type described above.  Consider for example a coupling
$g^{rl}=g_1$ for $r\le f_{R1}$ and $g^{rl}=g_2$ for $r>f_{R1}$.
This breaks the $SU(f_R)\times U(1)$ right flavor and charge symmetry
down to
$SU(f_{R1})\times U(1)\times SU(f_{R2})\times U(1)$ with
$f_{R1}+f_{R2}=f_R$.
Clearly we want to bosonize the two groups of right fermions separately,
introducing spin densities $S^i_{R1}$ and $S^i_{R2}$ generating $SU(2)$
Kac-Moody algebras of level $f_{R1}$ and $f_{R2}$, so that
the interaction will again involve only the spin sector of the theory.

In the limit $g_1 \gg g_2$, we can regard the $g_1$ interaction as
generating
precisely the flow described above, approaching arbitrarily closely to
the IR
fixed point described above.  We can then identify the $g_2$ interaction
as a
specific perturbation of this IR fixed point using our earlier results.
If it is still marginally relevant, this will produce a flow to a
final IR fixed point.
Of course this analysis would be reversed for $g_2 \ll g_1$.
For the intermediate regime, we can make a guess as to the likely behavior
by appealing to the $c$-theorem:
if one of the two final IR fixed points reached by the two limits of
extreme anisotropy has higher $c$, it is likely that any
finite anisotropy will cause the flow to continue to the other IR fixed
point.

There are several patterns which can arise in our example.
If $f_{R1}\ge f_L\ge f_{R2}$, the $g_1 \gg g_2$ limit will start with the
flow $
SU(2)_{f_L}\otimes SU(2)_{f_{R1}}\times SU(2)_{f_{R2}} \rightarrow
{SU(2)_{f_L}\times SU(2)_{f_{R1}-f_L}\over SU(2)_{f_{R1}}} \otimes
SU(2)_{f_{R1}-{f_L}}\times SU(2)_{f_{R2}}. $
The remaining interaction is irrelevant at this fixed point.
The $g_2 \gg g_1$ limit will follow a different  sequence:
$
SU(2)_{f_L}\otimes SU(2)_{f_{R2}}\times SU(2)_{f_{R1}}
\rightarrow
SU(2)_{{f_L}-f_{R2}}\otimes
{SU(2)_{f_{R2}}\times SU(2)_{{f_L}-f_{R2}}\over SU(2)_{f_L}}
\times SU(2)_{f_{R1}} \rightarrow
{SU(2)_{{f_L}-f_{R2}}\times SU(2)_{f_{R1}+f_{R2}-{f_L}}\over
SU(2)_{f_{R1}}}\otimes
{SU(2)_{f_{R2}}\times SU(2)_{{f_L}-f_{R2}}\over SU(2)_{f_L}}
\times SU(2)_{f_{R1}+f_{R2}-{f_L}}.$
If $f_L\ge f_{R1}\ge f_{R2}$, the two limiting sequences are both of
the latter form
 -- precisely this if $g_2 \gg g_1$, and with $f_{R1}$ and
$f_{R2}$ interchanged for $g_1 \gg g_2$.
One can check that in either case, if $f_{R1}> f_{R2}$,
the
result of the $g_2 \gg g_1$ sequence always has lower $c$ than the
result of the
$g_1 \gg g_2$ sequence, making it the IR fixed point for generic
anisotropy. The correlation functions for this case can be
found by the same means and will be given in a subsequent work.

One may also drive the charge sector to  universal
NFL fixed points. Indeed, the charge sector of $H_0$ has a 
 particle-hole $SU(2)$ symmetry
with the (right) charge being the third component $C_{z,R,r}={1\over2}\int
 \psi^*_{R,a,r}(x) \psi_{R,a,r}(x)$, $C^+_{R,r}=\int  \psi^*_{R,\uparrow,r}(x)
\psi_{R,\downarrow,r}^*(x)$ and $C^-=(C^+)^*$. Then, the previous
arguments indicate that adding an interaction term (an umklapp term) $
\sum  \int dx \; (g_c)^{rl}C^i_{R,r}(x) C^i_{L,l}(x)\; $ would
drive the model to a chiral NFL fixed point in the charge sector.
Again, the resulting charge correlation functions would be characterized by
the exponents discussed above.

\bigskip

\begin{figure}[here]
\psfig{file=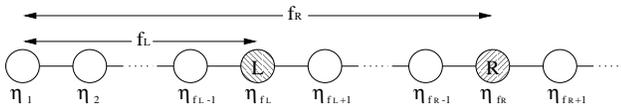,width=3.250in}
\bigskip
\caption{Diagrammatic representation of the TBA equations. Each circle
corresponds to a function $\eta_j$, and the links indicate which
functions are connected by the equations. The equations for $\ln
\eta_{f_L}$ and $\ln \eta_{f_R}$ have driving terms, which are
represented by filled circles.}
\label{fig3}
\end{figure}

\begin{figure}[here]
\psfig{file=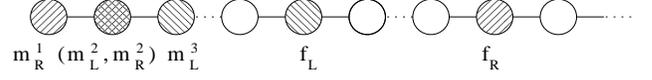,width=3.250in}
\bigskip
\caption{Same as the previous figure, but for a system with channel
anisotropy}
\label{fig3b}
\end{figure}

We can actually follow the flow at any scale by solving the model
exactly.
 The model exhibits dynamical fusion \cite{ad}
 (a different approach
was given in \cite{pw}), and allows a solution
  by a method developed in the context of the
anisotropic multichannel Kondo model \cite{pap}.
We find that
the model generates scales $m^l_L,m^r_R: \; l\le f_L,~ r\le f_R$,
parameterizing the patterns of
flavor symmetry breaking, and setting the scales of the 
excitation energies and
momenta. The free energy in the spin sector, the only sector
 modified by
the interaction with the impurity,  
 is given by,
\begin{eqnarray}
F(T,h) &=&
-\frac{TL}{2\pi}\int_{-\infty}^{\infty} d\xi \left(\sum_r  m^r_R e^{-\xi}
\ln(1+\eta_{r}(\xi,\frac{h}{T})) \right. \nonumber \\ && \left. + \sum_l  
m^l_L e^\xi \ln(1+\eta_{l}(\xi,\frac{h}{T}))\right), 
\label{freeen}
\end{eqnarray}
where the functions
$\{\eta(\xi,\frac{h}{T})\}$ are the solution of the following
system of coupled integral equations (TBA-equations, see Figs. 
\ref{fig3}-\ref{fig3b}):

\begin{eqnarray}
\ln \eta_n &=& -2\frac{m^n_L}{T} e^\xi  -2\frac{m^n_R}{T}
e^{-\xi}  + G\ln(1+\eta_{n+1}) \nonumber \\ &&  +
G\ln(1+\eta_{n-1}),~~~n=1,...,\infty,~~\eta_0\equiv 0,
\label{tbaeq}
\end{eqnarray}
with boundary condition:
$\ln \eta_{n} \rightarrow 2n\mu h/T$. The integral operator
 $G$ is defined by the kernel
$1/(2\pi\cosh(\xi'-\xi))$. In the isotropic case, $m_L^l=m_L
\delta_{f_L,l}, \; m_R^r=m_R \delta_{f_R,r}$.

The forms of the
driving terms $m_L e^\xi$ and $m_R e^{-\xi}$ are characteristic of
 massless left and
right moving excitations, and when both occur at the same level
(namely, when $f_L=f_R$)
 a driving term $m \cosh \xi$ results, indicating a mass gap.
To be more explicit, consider the
case of  two-channels of  right movers and one-channel of left movers.
The two
 couplings $g_1,~g_2$, are obtained by diagonalizing
the matrix of couplings $g_s^{rl}$. Choose  $g_1 < g_2, \;\phi = g_1/g_2$.
The physical scales then are, $
m^1_R = 2D_R \cos (\frac{\pi \phi}{2}) e^{-\frac{\pi}{g_2}}; \; m_R^2 =
D_R e^{-\frac{\pi}{g_1}} \;$ and $
m_L^1 = 2D_L  e^{-\frac{\pi}{g_1}}
$, with $D_L, D_R$ the densities of left and right movers \cite{pap}.

 From the  TBA-equations both the IR and the UV limits can be read off using
TBA-rules \cite{ken}: the IR limit of the left movers is obtained from
 the equations by
considering the right-mass as infinitely heavy and
truncating the equations  at the level it was inserted. An analogous
rule holds
for the right movers. The UV limit is obtained, on the other hand, by
considering the masses as vanishing. Applying these rules we deduce
the IR limit, and find accord with the conformal considerations.
The solution of the equations provides in addition the full
interpolation between the UV and IR limits 

\bigskip

\begin{figure}[here]
\psfig{file=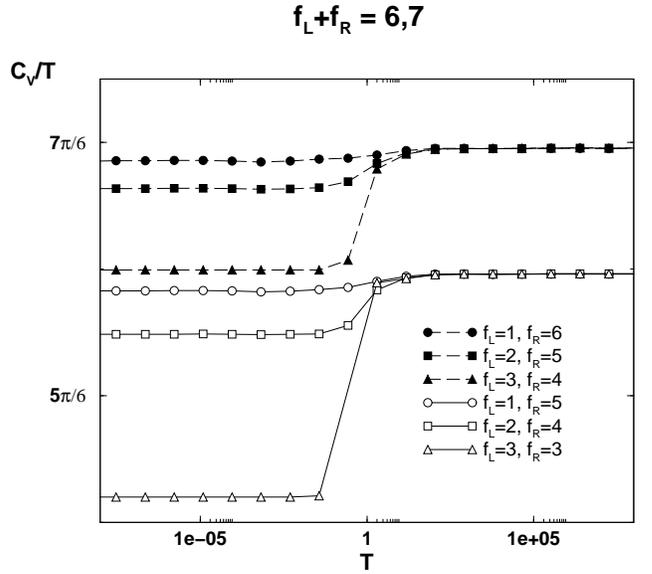,width=3.250in}
\bigskip
\caption{Numerical Solution of the TBA equations: $C_V/T$ vs. $T$ for
systems with $f_L+f_R=6,7$}
\label{fig4}
\end{figure}

\begin{figure}[here]
\psfig{file=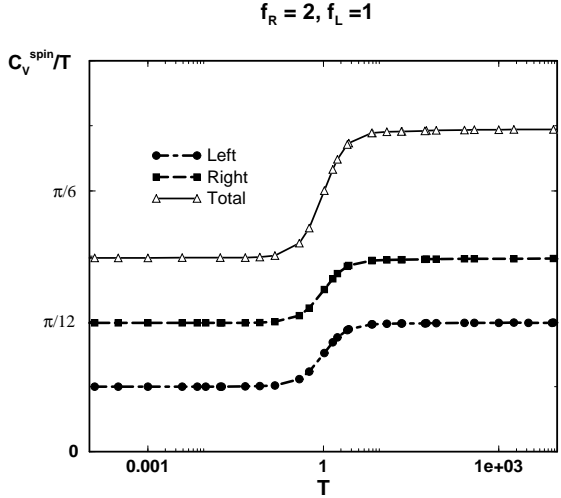,width=3.250in}
\bigskip
\caption{Spin part of $C_V/T$ for $f_R=2, f_L=1$. The lower curves
correspond to the right and left movers. }
\label{fig5}
\end{figure}

\begin{figure}[here]
\psfig{file=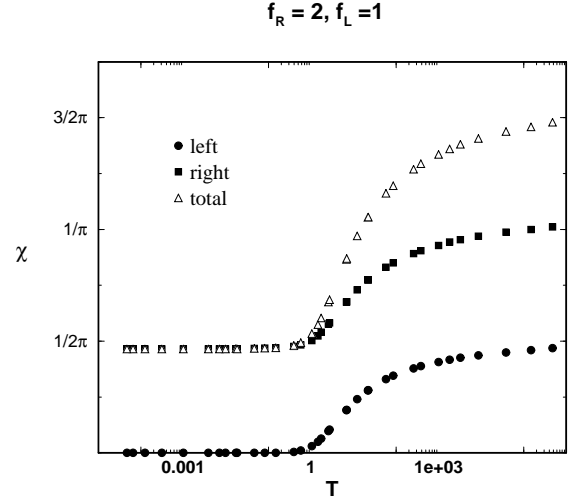,width=3.250in}
\bigskip
\caption{Magnetic susceptibility $\chi$ vs. $T$ for $f_R=2, f_L=1$.}
\label{fig6}
\end{figure}

\begin{figure}[here]
\psfig{file=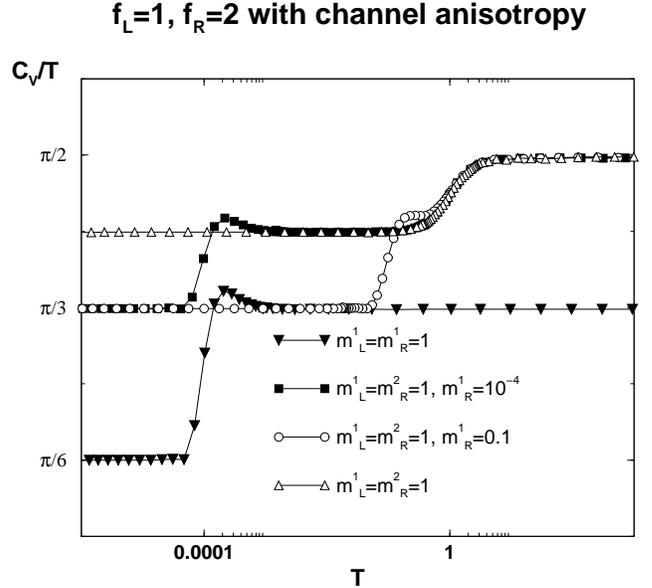,width=3.250in}
\bigskip
\caption{$C_V/T$ vs. $T$ for $f_R+f_L=2,3$ and different values of
Channel anisotropy. The legend indicates the values of $m$ which are
not zero. The isotropic cases correspond to the cases with only two
nonvanishing $m$: $m_L^1=m_R^2=1$, and $m_L^1=m_R^1=1$.} 
\label{fig7}
\end{figure}

\begin{figure}[here]
\psfig{file=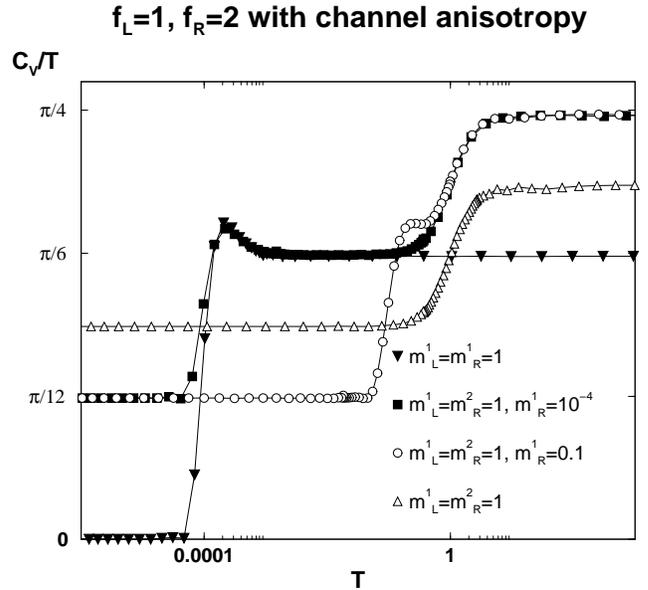,width=3.250in}
\bigskip
\caption{Same as the previous figure, but with only the contribution
of the spin degrees of freedom. } 
\label{fig8}
\end{figure}

\begin{figure}[here]
\psfig{file=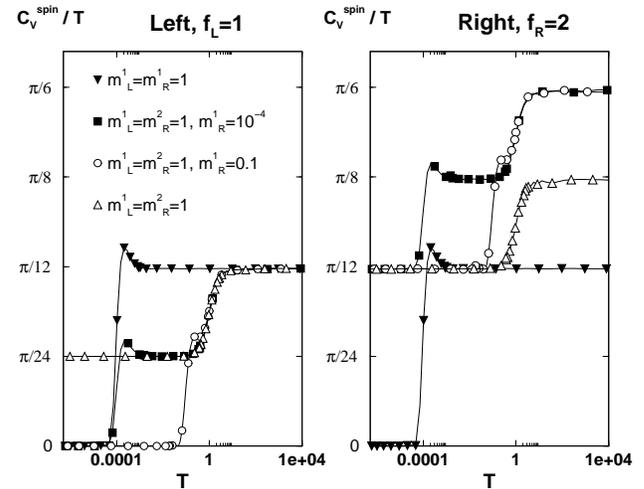,width=3.250in}
\bigskip
\caption{Spin part of $C_V/T$ vs. $T$, as in the previous figure,
separated into the contributions from left and right movers, 
respectively.} 
\label{fig9}
\end{figure}

We have solved the TBA equations (\ref{freeen},\ref{tbaeq}) 
numerically 
for $f_R+f_L=2,3,6,7$, and for several values of $f_R-f_L$. The results 
are contained in Figs. \ref{fig4}-\ref{fig9}. 

Fig. \ref{fig4} 
corresponds 
to the linear coefficient of the specific heat for $f_R+f_L=6,7$. 
Since the TBA equations describe only the spin
sector, we have added to each curve the charge and flavor contributions.
This is a temperature independent term that can be read off eq.(\ref{central})
leading to 
$C_V/T=\frac{\pi}{6}(f_L+f_R)$  the noninteracting result at high-T. The 
temperature is expressed in energy units such that the energy  scales 
$m^r_{L,R}$ that appear in (\ref{freeen}) are set to unity.
The
crossover to a smaller value of $\gamma \equiv C_V/T$ at low-T reflects
the change in the spin contribution due to strong 
coupling. The values confirm the CFT prediction (\ref{estodo}), 
for $f_R \ne f_L$. For $f_R=f_L=3$, on the other hand, a spin gap
develops and the spin contribution falls precipitously, leaving only 
the  massless charge and flavor degrees of
freedom, unaltered by the interaction, to contribute to  $C_V/T$.

Within the Bethe Ansatz analysis it is possible to consider only the spin 
contribution to the specific heat as well as the individual contributions 
of the right and the left movers to the
thermodynamics. We have made such analysis in detail for the case 
$f_R=2, f_L=1$.
The spin sector in the UV regime correspond to $SU(2)_2 \otimes SU(2)_1$.
Therefore, the CFT analysis yields $\gamma_{spin}^{uv} = \frac{\pi}{12} 
\left(\frac{3}{2} +1 \right)$. The first term correspond to the right movers,
as confirmed by the high-T region of Fig. \ref{fig5}, whereas the second 
term corresponds to the left
movers, which are just spinors. 
The IR fixed point has the structure (\ref{eq7}), 
$\frac{SU(2)_1 \times SU(2)_1}{SU(2)_1} \otimes SU(2)_1$,
which yields $\gamma_{spin}^{ir} = \frac{\pi}{12} \left(1+\frac{1}{2}
\right)$ (see low-T part of Fig. \ref{fig5}). The spin 
part of the left movers is described by a Ising model (one majorana fermion) 
which has
central charge 1/2, while the right
movers are equivalent to regular spinons, with central charge 1. However, 
the right movers do not behave like regular electrons, since they have 
a $SU(2)_2$ flavor sector. Indeed, the total low-T linear
coefficient of the specific heat is $\gamma^{ir}= \frac{\pi}{12} \left(
\frac{7}{2} + \frac{3}{2} \right) = \frac{5\pi}{12} $, as is shown in 
Fig. \ref{fig8}. Hence both right and left movers have correlation
functions with anomalous exponents.

The calculation of the magnetic susceptibility, Fig. \ref{fig6}, 
indicates a crossover 
from a high-T regime where there are three kinds of spinons (two right movers
and one left movers) to a completely different low-T regime. In the IR 
fixed point, the left movers form a singlet, as can be seen in Fig. \ref{fig6}
by the
vanishing of $\chi^{left}$ at low-T, while the right movers contribution
corresponds to just one spinon. The Wilson ratio is $
R_W = \frac{\gamma^{ir}}{\chi^{ir}}\frac{3}{\pi^2} = \frac{5}{2}$
confirming the CFT prediction.

We consider also the effects of channel anisotropy in the $f_R=2$, 
$f_L=1$ case, Figs.\ref{fig7}-\ref{fig9}. Channel anisotropy in the 
right moving sector becomes manifest in the 
TBA equations (\ref{tbaeq}) by the appearance of additional driving 
terms at levels
below $f_R$ (equivalently for the left movers). For small
anisotropy (when $m_R^{r<f_R} \ll m_R^{f_R}$), the $T \ll m_R^{r<f_R}$ 
behavior of the
system corresponds to a new fixed point, characterized by the value of
$r$ (the level of the new driving term in TBA and in Fig. \ref{fig3b}). 
There is also
an intermediate regime, $m_R^{r<f_R} \ll T \ll m_R^{f_R}$, where the 
behavior of the system is similar to the isotropic fixed point.

The behavior of $\gamma$ is shown for
several values of channel anisotropy  in Fig. \ref{fig7}. We have also
included the isotropic case, and the $f_L=f_R=1$ case. At high temperatures, 
the behavior is the characteristic of the free model, with $\gamma^{uv} 
\equiv \frac{\pi}{6} (f_R+f_L)$. As the temperature is lowered, there is a
crossover to the strong coupling regime. If the anisotropy is small enough,
the system has an intermediate regime characterized by the same central
charge as the isotropic fixed point. As the anisotropy increases, the
intermediate region disappears. For low enough temperatures,
$T \ll m_R^1$, the system flows to a different fixed point, with the
same central charge as the high-T, $f_R=f_L=1$ case. Thus, even though the
flavor symmetry has been broken, there are still massless modes in the
spin sector. Notice that the low-T behavior of the  $f_L=f_R=1$ is 
characterized by the opening of a gap in the spin sector and a drop in 
the value of $C_V/T$. We also observe a peak in $C_V/T$ at the crossover region
for any anisotropy.
Fig. \ref{fig8} corresponds to the spin part of $C_V/T$, which has 
been divided into 
right and left contributions in Fig. \ref{fig9}. The first thing to notice in 
Fig. \ref{fig8} is that the value of $\gamma^{uv}_{spin}$ in the cases with 
anisotropy is
larger than the isotropic value. This is so because, as indicated in the
CFT analysis, one starts from different symmetries in the weak coupling
region. On the one hand the  isotropic case has a symmetry 
$(SU(2)_2 \times SU(2)_2 \times U(1))\otimes (SU(2)_1\times U(1))$ 
which yields $\gamma^{uv}_{spin} = \frac{\pi}{12}\left(\frac{3}{2}+1\right)$ 
at high-T. On the other hand, the anisotropic case has a symmetry
$(U(1)\times U(1) \times SU(2)_1 \times SU(2)_1)\otimes (SU(2)_1\times U(1))$.
This yields  $\gamma^{uv}_{spin} = \frac{\pi}{12}\left(2+1\right)$ at high-T. 
This
can also be seen in Fig. \ref{fig9}. For small anisotropy there is a 
crossover to
an intermediate regime where $\gamma^{intermediate}_{spin}=
\frac{\pi}{12}\left(\frac{3}{2}
+\frac{1}{2}\right)$ whereas for the isotropic case $\gamma^{ir}_{spin}= 
\frac{\pi}{12} \left( 1+\frac{1}{2} \right)$. Even though there is a difference
in the spin sector, the total values of $\gamma$ coincide, as we saw in Fig. 8.
Notice also that the total spin part in the intermediate regime 
coincide with the high-T value for $f_L=f_R=1$. However, in the latter 
case,
there is the same spinon contribution from right and left movers, whereas in 
the former case we have contributions from objects with central charge
$\frac{3}{2}, \frac{1}{2}$, respectively.

Finally, at low enough temperatures, the left modes become massive, as in the
$f_L=f_R=1$ case, and their contribution to $\gamma_{spin}$ vanishes. 
All the specific heat from spin modes comes from the right movers, with a 
contribution $\gamma_{spin}=\frac{\pi}{12}$,
corresponding to a single spinon mode.

\bigskip

 We return now to
the example of the paired QHE samples presented earlier. Virtual hopping
between the edges will generate interaction terms of the form
(\ref{eq}) which will drive
spin sector to
 exhibit a universal behavior characteristic of a chiral  liquid.
This fixed point will determine the behavior of the system down to
energy scales $\sim E_z$ below which the system will flow 
  to more conventional fixed points. 
The effect of the chiral fixed point 
 will show up in the spin susceptibilty, in the
spin conductance  $G_s$ or in the NMR response encoded in the spin
correlation
function
$<S^i(t,0)S^i(0,0)>=<S_L^i(t,0)S_L^i(0,0)>+<S_R^i(t,0)S_R^i(0,0)>$. From 
our previous results we thus have
\begin{eqnarray}
{1 \over TT_1}= {\chi^{``}_{ii} \over \omega} = constant + \omega
^{\alpha} \nonumber
\end {eqnarray}
where $\alpha= 4/(f_R-f_L +2)$, except  when $f_R-f_L=1$, in which case
$\alpha= 2$.

We have thus proposed and studied
models of interacting chiral fermions
which realize a new class of non-Fermi liquids in one dimension.
A general mechanism, chiral stabilization, was found to be the source
of this behavior.  It would apply in any model of chiral fermions with
non-abelian symmetry, and in some cases even with weakly broken
non-abelian symmetry.

The generality of this mechanism suggests
that many other experimental realizations will be found and also new
fixed points will be identified. Higher hierarchy edge states in QHE
systems furnish an example for both. When the  filling factor is
$\nu=n/(np+1)$
with $p$ a negative even number  one edge state mode moves in a
direction opposite  to the rest \cite{wfi} providing a chiral
imbalance. These modes are no longer fermions but chiral luttinger 
excitations and when allowed to interact in the manner discussed above
new 
fixed
points are expected to arise which would be observable in very clean samples.
\bigskip

Acknowledgments: It is a pleasure to thank E. Y. Andrei, P. Coleman, K.
Intrilligator, L.
Ioffe, J. Jain, A. Lopez, A.
Ruckenstein and A. J. Schofield for comments, suggestions
and enlightening discussions. AJ was supported by EPSRC grant GR/K97783.


\begin{references}
\bibitem{corr}  The leading
correction (in the infinite cut-off limit)
 to the linear term in the specific heat
 is $
 T^{2+4/(f_R-f_L+2)}$. It comes from the leading singlet
 irrelevant operator at the IR fixed point
$\bar\phi^{0,1}_{0} J^a g^{1a}$.
\end{references}
\end{document}